\documentclass[12pt]{iopart}

\usepackage{iopams}

\usepackage{graphicx}
\usepackage{dcolumn}
\usepackage{bm}

\newcommand{\Ji}{\mathit J_{\mathrm{sd}}}
\newcommand{\rv}{\bi r}

\begin{document}

\title{Influence of the Coulomb interaction on the exchange coupling in granular magnets}

\author{O.~G.~Udalov}
\address{Department of Physics and Astronomy, California State University Northridge, Northridge, CA 91330, USA}
\address{Institute for Physics of Microstructures, Russian Academy of Science, Nizhny Novgorod, 603950, Russia}
\ead{oleg.udalov@csun.edu}

\author{I.~S.~Beloborodov}
\address{Department of Physics and Astronomy, California State University Northridge, Northridge, CA 91330, USA}


\date{\today}

\begin{abstract}
We develop a theory of the exchange interaction between ferromagnetic (FM) metallic grains embedded into insulating matrix
by taking into account the Coulomb blockade effects.
For bulk ferromagnets separated by the insulating layer the exchange interaction strongly depends on the
height and thickness of the tunneling barrier created by the insulator.
We show that for FM grains embedded into insulating matrix
the exchange coupling additionally depends on the dielectric properties of this matrix
due to the Coulomb blockade effects. In particular, the FM coupling
decreases with decreasing the dielectric permittivity of insulating matrix.
We find that the change in the exchange interaction due to the Coulomb blockade effects can be a few
tens of percent. Also, we study dependence of the intergrain exchange interaction on the grain size and
other parameters of the system.
\end{abstract}

\submitto{\JPCM}

\noindent{\it magneto-electric effect, multiferroics, Coulomb blockade, granular}:

\maketitle

\section{Introduction}

Physics of granular ferromagnets (GFM) combines numerous phenomena appearing at different length and energy scales. This makes GFM a complicated object suitable for investigation
of fundamental effects and their mutual influence~\cite{Sellmyer2016, Nogues2015, Nolting2014,Zheng2015, Fullerton2014, Gerber2014, Carman2013, Woinska13, Bayer2011, Bertram2001, Bel2014ME, Bel2014ME1, Bel2014ME2, Bel2014ME3}.
Disorder combined with strong Coulomb interaction leads to peculiar dependence of the conductivity of
granular metals on temperature~\cite{Bel2007review}. The Coulomb blockade strongly affects superconductivity in granular superconductors~\cite{Deut1983,Vill2004,Efetov2002}. Granular ferroelectrics demonstrate the
metal-insulator transition driven by temperature and electric field~\cite{Bel2014GFE1,Bel2014GFE2,Bel2013GFE}. In granular materials with small grains the size quantization effects become important~\cite{Bel2007review}. Besides, these materials are known as good candidates for various applications~\cite{Hadjipanayis1999,Masumoto2007,Masumoto2006}.

In this manuscript we study magnetic properties of granular ferromagnets - materials with ferromagnetic
(FM) metallic grains embedded into insulating matrix.
The magneto-dipole (MD)~\cite{Ayton95,Ravichandran96,Djurberg97,Sahoo03} and the exchange interactions~\cite{Freitas2001,Sobolev2012, Hembree1996, Lutz1998,Freitas2007} are the main intergrain coupling mechanisms in GFM. Long range MD interaction leads to the formation of super spin glass (SSG) state in the system. Depending on its sign the short range exchange coupling causes the formation of either super ferromagnetic (SFM) state or
the SSG state.
In addition to the interparticle interaction the magnetic anisotropy of individual grains influences
the properties of the GFM leading to ``blocking'' phenomena~\cite{Chien1988,Chien1991, Kechrakos98, Grady1998}.
\begin{figure}
\includegraphics[width=0.5\columnwidth]{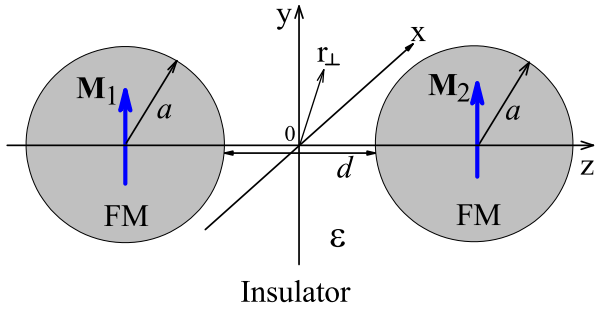}
\caption{(Color online) Two FM metallic grains with radius $a$ and intergrain distance $d$ embedded
into insulating matrix with dielectric constant $\varepsilon$. $\bi M_{1,2}$ stands for grain
magnetic moment. (x,y,z) is the coordinate system. $\bi r_\perp$ is the radius vector in the (x,y) plane. } \label{Fig:Model}
\end{figure}

For well separated grains the interparticle coupling is weak and the
properties of GFM are defined by the single particle magnetic anisotropy.
This situation is well studied theoretically and experimentally.
As the grains move closer to each other the MD interaction becomes important.
For even smaller distances, of the order of 1 nm, the exchange coupling becomes crucial.
The influence of the exchange interaction on macroscopic
magnetic state of GFM is understood using the Heisenberg model, $\sum_{ij}J_{ij}(\bi M_i\bi M_j)$~\cite{Hembree1996,Ziemann2004,Grady1998}, where summation is over the
all nearest neighbor grain pairs in the whole GFM, $\bi M_i$ is the magnetic moment
of grain $i$ and $J_{ij}$ is the exchange coupling constant for grain pair \{$ij$\}. However,
the microscopic theory of the exchange interaction between magnetic grains (constant $J_{ij}$) is still lacking.
The understanding of the intergrain exchange interaction is mostly based on the Slonczewski theory
developed for coupling of infinite FM layers separated by the insulating layer~\cite{Slonczewski1989,Bruno1995}.
According to Slonczewski the interlayer exchange coupling appears due to virtual electron hopping (or tunnelling) between FM leads.
This theory does not take into account the many-body effects and charge quantization phenomena.
The influence of many-body effects on the ground magnetic state and transport properties of
solid state systems is a long standing fundamental problem appearing in a broad range of physical problems.
In nanoscale granular systems the many-body effects due to Coulomb interaction
between electrons become crucial~\cite{Bel2007review}.
In particular, these effects influence the electron transport in granular metals~\cite{Bel2007review}, ferromagnets~\cite{Bel2007},  superconductors~\cite{Adams1994} and ferroelectrics~\cite{Bel2014GFE1,Bel2014GFE2}.
In this paper we show another example of importance of many-body effects in granular systems.
We develop a theory of the exchange interaction between FM nanograins embedded into insulating matrix
by taking into account the Coulomb blockade effects (see figure~\ref{Fig:Model}) and show that in granular systems
the matrix dielectric constant and the grain size influence the intergrain exchange 
interaction. Note that the typical grain sizes considered in this paper is in the range of few nms 
with thousands of atoms. The Coulomb blockade is important for such grains. At the same time 
the grains can be treated as bulk metal neglecting surface and size quantization effects such as in magnetic clusters made of several atoms.

Note that experimental realization of the GFM with the intergrain
exchange coupling meets a number of difficulties. The most complicated
task is the control of the intergrain distance on the scale of 1 nm. Nonetheless, several studies observed
the intergrain exchange interaction of FM type with rather large value~\cite{Freitas2001,Sobolev2012,Hembree1996}.

The paper is organized as follows.  We summarize our main results in Sec.~\ref{Results}.
We introduce the model to study the exchange interaction between two magnetic grains in Sec.~\ref{Sec:model}.
In Sec.~\ref{grainfunctions} and \ref{Exchange}
we calculate the exchange interaction between two ferromagnetic grains embedded into
insulating matrix. In Sec.~\ref{Sec:Anal} we analyze the dependence of the intergrain
exchange interaction on the system parameters. We discuss validity of our model in Sec.~\ref{Sec:Val}.

\section{Main results}  \label{Results}

Here we summarize our main findings.

1) We develop a theory of the exchange interaction between FM metallic grains embedded into insulating matrix
by taking into account the Coulomb blockade effects.
We show that beside height and thickness of the insulating barrier the intergrain exchange coupling depends on the dielectric properties of the spacer. This additional dependence occurs
due to the Coulomb blockade effects (see figure~\ref{Fig:JvsJ1}).
The FM coupling decreases with decreasing the dielectric permittivity of insulating matrix.

2) We predict the behavior of intergrain exchange interaction $J$
as a function of grain size $a$. On one hand increasing the grain size
leads to the linear increase of the interaction due to geometrical factor, $J\sim a$.
On the other hand decreasing the grain size leads to the enhancing of
the Coulomb blockade and decreasing the FM contribution to the exchange coupling.
Finally, the exchange interaction decays faster than the first power
of $a$ with decreasing the grain size.

3) We find that the Coulomb blockade influences the intergrain exchange coupling strongly if:
i) the Fermi momentum is not too large and ii) the spin subband splitting is of
the order of the Fermi energy (see figure~\ref{Fig:Peak}).

4) We show that the impact of the Coulomb blockade also depends on
the barrier thickness and height: The thicker and the lower the barrier -
the stronger the Coulomb blockade effect.

\section{The model}\label{Sec:model}

\begin{figure}
\includegraphics[width=0.5\columnwidth]{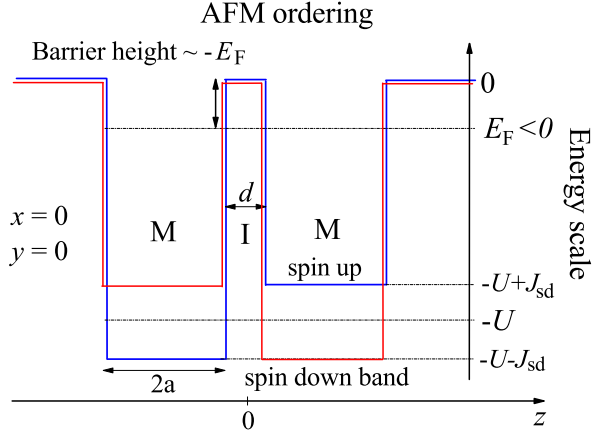}
\caption{(Color online) Schematic picture of potential energy profiles
for electron with spin ``up'' (red line) and spin ``down'' (blue line)
states for AFM configuration of grain magnetic moments. Profiles are slightly shifted
with respect to each other for better viewing. Zero energy corresponds to
the top of energy barrier for electrons in the insulator.
Symbols M and I stand for metal and insulator, respectively. All other notations are defined
in the text.}  \label{Fig:EnScales}
\end{figure}

We consider two identical FM grains with radius $a$ and the distance between the grain surfaces
being $d$ (see figure~\ref{Fig:Model}). The Hamiltonian describing delocalized electrons in the system has the form
\begin{equation}\label{Eq:HamIn}
\hat H=\hat H_\mathrm{sp}+\hat H_{\mathrm{C}},
\end{equation}
where single particle Hamiltonian $\hat H_{\mathrm{sp}}=\sum_{i}(\hat K(\rv_i)+\hat U_1(\rv_i)+\hat U_2(\rv_i)+\hat H_{1\mathrm{m}}(\rv_i))+\hat H_{2\mathrm{m}}(\rv_i))$ consists of kinetic energy $\hat K$, potential energy $\hat U_{1,2}$ and exchange interaction between delocalized and localized electrons, $\hat H_{1,2\mathrm{m}}$. Summation is over $2n_0$ electrons in the system, where $n_0$ is the number of electrons in each grain. The single particle potential energy inside the grain (1) is $\hat U_1=-U$ ($U>0$) and $\hat U_1=0$ outside the grain (1), $\hat U_2=-U$ inside the grain (2) and $\hat U_2=0$ outside the grain (2).
We consider FM and AFM configurations of grain magnetizations.
Therefore magnetic interaction has the form,
$\hat H_{1,2\mathrm{m}}=-J_\mathrm{sd}\hat \sigma_z M_{1,2}$ inside the grains,
with  $M_{1,2} = \pm 1$, $\hat \sigma_z$ being the Pauli matrix and
$\Ji$ being the coupling constant of the
s-d interaction responsible for spin subband splitting of conduction electrons.
The energy profiles for spin up and spin down subbands are shown in figure~\ref{Fig:EnScales} for AFM configuration of grain magnetic moments.

We introduce a single particle Hamiltonian for each grain,
$\hat H^{\mathrm{g}}_{1,2}=\hat K+\hat U_{1,2}+\hat H_{1,2\mathrm{m}}$, with
eigenfunctions $\psi_i^s$ and $\phi_i^s$ for grain (1) and (2), respectively. Here we note that the
single particle Hamiltonian of two grains is not the sum of $\hat H^{\mathrm{g}}_{1}$ and $\hat H^{\mathrm{g}}_{2}$.
The subscript $i$ stands for orbital state and the superscript $s$ denotes the spin state in a local spin coordinate system related to magnetization of corresponding grain.
Due to grains symmetry the wave functions are symmetric
\begin{equation}\label{WFSym}
\psi^s_i(x,y,z)=\phi^s_i(x,y,-z).
\end{equation}
Since we consider identical grains, energies of these states are
equal and denoted $\epsilon_i^s$.

Functions $\psi_i^s$ are orthogonal to each other and normalized,
$\langle\psi_i^{s}|\psi_j^{s'}\rangle=\delta_{i,j}\delta_{s,s'}$ and
$\langle\phi_i^{s}|\phi_j^{s'}\rangle=\delta_{i,j}\delta_{s,s'}$. However,
functions $\psi_i^s$ and $\phi_{i}^s$ are not orthogonal to each other,
$\langle\phi_i^{s'}|\psi_j^{s}\rangle=P^{s}_{ij}\delta_{s,\pm s'}$ (symbol $+$ stands for FM and $-$ is for AFM configurations).
We assume that the barrier between the grains is high enough such
that the wave functions overlap integral, $P^s_{ij}$ is small, $P^s_{ij}\ll 1$.
The wave functions $\psi$ and $\phi$ exponentially decay outside the grains with
some characteristic length scale $1/\varkappa$ ($\psi\sim \rme^{-\varkappa r_1}$ and $\phi\sim \rme^{-\varkappa r_2}$,
where $r_1 (r_2)$ is the distance from the centre of the corresponding grain).
Due to the exponential decay the overlap is small
and can be estimated as $P^s_{ij}\sim \rme^{-\varkappa d}$. At the Fermi
level $E_\mathrm F$ (in our consideration $E_\mathrm F < 0$, see figure~\ref{Fig:EnScales})
the inverse decay length is $\varkappa_0\approx\sqrt{-2m_\mathrm eE_\mathrm F/\hbar^2}$.
Thus, increasing the barrier thickness or the barrier height one can control the smallness of
the overlap integral. Below we use, $p=\rme^{-\varkappa_0 d}$, as the small parameter in the problem.
We neglect all states with energies $\epsilon^s_i>0$ since these states are fully delocalized.

The zero-order many-particle wave function $\Psi_0$ corresponds to the system state with all
single particle states $\psi^s_i$ and $\phi^s_i$ with energies $\epsilon^s_i<E_\mathrm F$ being filled and with all states
above $E_\mathrm F$ being empty (see Appendix~\ref{Formalism} for details).  The creation and annihilation
operators are $\hat a^{s+}_i$ and $\hat a^{s}_i$ in grain (1), and
$\hat b^{s+}_i$ and $\hat b^{s}_i$ in grain (2). We introduce here the excited wave functions $\Psi_{ij}^s=\hat a^{s'+}_{i}\hat b^{s}_{j}\Psi_0$ and $\tilde \Psi_{ij}^s=\hat b^{s'+}_{i}\hat a^s_{j}\Psi_0$ ($s'=s$ for FM orientation of $\bi M_1$ and $\bi M_2$ and $s'=-s$ for AFM configuration). We neglect states with two electrons being transferred
between grains since these states have much larger Coulomb energy.

We use the simplest model for Coulomb interaction with diagonal elements only.
The zero order wave function corresponds to the system state
where both grains are neutral and the Coulomb energy is zero,
$\langle\Psi_0|\hat H_\mathrm C|\Psi_0\rangle=0$. In the
excited states $\Psi^s_{ij}$ and $\tilde \Psi^s_{ij}$ an electron is transferred from one grain into
another. Therefore, the grains have opposite charges and the
energy of the Coulomb interaction is $\langle\Psi_{ij}^s|\hat H_\mathrm C|\Psi_{ij}^s\rangle=\langle\tilde\Psi_{ij}^s|\hat H_\mathrm C|\tilde\Psi_{ij}^s\rangle=\tilde\epsilon_\mathrm c=e^2(C^{-1}-(2C_\mathrm m)^{-1})$. This is just the
classical electrostatic energy of two oppositely charged metallic spheres.
Here $C$ is the single grain capacitance and $C_\mathrm m$ is the mutual capacitance of
two grains. We can estimate the charging energy as $\tilde \epsilon_\mathrm c=e^2/(8\pi a\varepsilon\varepsilon_0)$ for $d\approx1$ nm and $a\in\left[1;10\right]$ nm, here $\varepsilon$ is the effective dielectric permittivity of GFM and $\varepsilon_0$ is the vacuum dielectric constant. We assume that the Coulomb interaction does not transfer
electrons between grains, or the Coulomb-based hopping is negligible in comparison with
the hopping due to kinetic energy. Thus, we have
$\langle\Psi_{ij}^s|\hat H_\mathrm C|\Psi_0\rangle=\langle\tilde\Psi_{ij}^s|\hat H_\mathrm C|\tilde\Psi_0\rangle=\langle\Psi_0|\hat H_\mathrm C|\Psi_{ij}^s\rangle=\langle\tilde\Psi_0|\hat H_\mathrm C|\tilde\Psi_{ij}^s\rangle=0.$ This model
for the Coulomb interaction is valid for metallic grains with
large conductance~\cite{Bel2007review,Glazman2002}.

We will study two cases: 1) FM and 2) AFM alignment of grain magnetic moments.
Both these configurations are collinear meaning that the single particle
interactions $\hat W_\mathrm k$, $\hat U_{1,2}$ and $\hat H_{1,2\mathrm{mag}}$ are diagonal in the spin space.

Below we will find the energy of the system for FM ($E^{\mathrm{FM}}$) and AFM ($E^{\mathrm{AFM}}$)
alignment and calculate the intergrain magnetic (exchange) interaction,
\begin{equation}\label{Eq:ExFin}
J=E^{\mathrm{AFM}}-E^{\mathrm{FM}}.
\end{equation}
For $J >0$ the interaction between the grains is FM
while for $J < 0$ it is AFM. We consider the case of zero temperature and
therefore neglect all inelastic transitions of electrons between
grains due electron-phonon interaction. Thus, we take into account only co-tunneling
processes neglecting sequential tunneling (see Sec.~\ref{Sec:Val} for more details).

The Hamiltonian in~(\ref{Eq:HamIn}) does
not include the vector potential occurring due to magnetic field produced
by the grains. This contribution is small in comparison with s-d exchange coupling and
can be neglected. Also we neglect the MD interaction. On one hand this interaction
was considered in numerous papers in the past. On the other hand the MD interaction is the long
range interaction and thus its consideration for two grains only is meaningless.
This coupling should be considered on the scale of whole granular magnet.

\section{Single grain wave functions and matrix elements}
\label{grainfunctions}

We use the following approximate wave functions to calculate all
matrix elements. Outside the grain the wave function corresponding to the
wave vector $\bi k$ and the spin state $s$ is
\begin{eqnarray}\label{Eq:WaveFuncR5}
\eqalign{\psi_\bi{k}^s(z,r_\perp)\approx\frac{\tau_\bi k^s}{\sqrt{\Omega}}\exp\left(-\varkappa_\bi k^s\left(\frac{d}{2}+ z+\frac{r_\perp^2}{2a}\right)\right)\rme^{\rmi \bi k_\perp \bi r_\perp},\cr
\phi_\bi{k}^s(z,r_\perp)\approx\frac{\tau^s_\bi k}{\sqrt{\Omega}}\exp\left(-\varkappa_\bi k^s\left(\frac{d}{2}
- z+\frac{r_\perp^2}{2a}\right)\right)\rme^{\rmi \bi k_\perp \bi r_\perp}.}
\end{eqnarray}
Here $\tau_\bi{k}^s =\frac{2k_z}{k_z+\rmi \varkappa^s_\bi k}$ is the amplitude of the transmitted electron wave,
$\bi k_{\perp}=(k_x,k_y,0)$, $\bi r_{\perp}=(x,y,0)$, $\Omega=4\pi a^3/3$ is the grain volume and
$\varkappa^s_\bi k=\sqrt{2m_\mathrm e(U-s\Ji-\hbar^2k^2_z/(2m_\mathrm e))/\hbar^2}$
is the inverse decay length.
We introduce the following coordinates: $z$ is along the line connecting grain centres;
$z=0$ is the symmetry point between the grains; $x$ and $y$ are perpendicular to $z$.
For electron wave function inside the grains we have
\begin{equation}\label{Eq:WaveFuncR6}
\eqalign{
\psi^s_\bi k(z,r_\perp)\approx\frac{\rme^{\rmi k_{z}\left(\frac{d}{2}+ z+\frac{r_\perp^2}{2a}\right)}+\xi^s_\bi k \rme^{-\rmi k_z\left(\frac{d}{2}+ z+\frac{r_\perp^2}{2a}\right)}}{\sqrt{\Omega}}\rme^{\rmi \bi k_\perp \bi r_\perp},\cr
\phi^s_\bi k(z,r_\perp)\approx\frac{\rme^{\rmi k_{z}\left(\frac{d}{2}- z+\frac{r_\perp^2}{2a}\right)}+\xi^s_\bi k \rme^{-\rmi k_z\left(\frac{d}{2}- z+\frac{r_\perp^2}{2a}\right)}}{\sqrt{\Omega}}\rme^{\rmi \bi k_\perp \bi r_\perp},
}
\end{equation}
with $\xi^s_{\bi k} = \frac{k_z-\rmi \varkappa^s_{\bi k }}{k_z+\rmi \varkappa^s_{\bi k }}$.

Tunneling matrix elements calculated in the following sections 
depend on the overlap of the electron wave functions located at different grains. 
The overlap region in the ($x,y$) plane is defined by $r_\perp<\sqrt{2a/\varkappa}$. 
This is the area where the wave functions are essentially non-zero. The estimate 
of the radius of the overlap region is correct if the grain size $a$ exceeds 
the intergrain distance $d$ and $r_\perp\ll a$ ($\varkappa a\gg 2$). 
The limit $d\ll a$ is valid since the exchange interaction decays fast 
with the intergrain distance.  We assume that the distance $d$ is of order of 1 nm, 
while the grain size is bigger. The second condition is satisfied even 
for 1 nm radius grains ($\varkappa=4$ nm$^{-1}$ for the barrier height of 0.5 eV). 
The effective contact area of the grains can be introduced as $S_\mathrm c=\pi a/\varkappa_0$. 
Within the contact area we will change the wave function with the plane waves 
neglecting the factor $\rme^{-\varkappa r^2_\perp/(2a)}$. Outside the contact 
region we will neglect the wave function overlap. The details are shown in Appendix A.

Below to simplify the notations we will use the
subscript $i$ (or $j$) to enumerate the electron states instead of wave vector $\bi k$.
Also we introduce here the area $S^{ij}_\mathrm c=2\pi a/(\varkappa_i+\varkappa_j)$ and the
corresponding length, $\lambda^{ij}_{\perp}=\sqrt{S^{ij}_\mathrm c/\pi}$. The more detailed
discussion of the wave functions is presented in \ref{grainfunctionsApp}.

We calculate the matrix elements of the single particle Hamiltonian $\hat H^\mathrm{sp}$ using
the wave functions $\psi$ and $\phi$ of isolated grains. We study separately the FM and AFM
configurations, since for AFM configuration the wave functions with the same
$z$-spin projection are different for different grains.

We introduce the following notations: $S^{\pm}$ is the set of single particle states with
spin being co-directed (``+'') and counter-directed (``-'') with grain
magnetization; $S_0^{\pm}$ denotes the subset of $S^{\pm}$ for states with $\epsilon^{\pm}_i<E_\mathrm F$.
These sets are identical for both grains.

To calculate the energy of two grains we use the following matrix elements
\begin{equation}\label{Eq:ME1}
\eqalign{
 P_{ij}^s=\langle\phi_i^{s'}|\psi_j^{s}\rangle,\\
 V^s_{ii}=\langle\psi_i^s|\hat U^\mathrm{sp}_2+\hat H_{2\mathrm{mag}}^\mathrm{sp}|\psi_i^{s}\rangle,\cr
 T^s_{ij}=\langle\phi_i^{s'}|\hat U^\mathrm{sp}_2+\hat H_{2\mathrm{mag}}^\mathrm{sp}|\psi_j^{s}\rangle.\cr
}
\end{equation}
Here $i\in S^s, j\in S^{s'}$, $s'=s$ for FM configuration and $s'=-s$ for AFM configuration.
All other matrix elements contributing to the energy of the system are small and can be neglected.
Using (\ref{Eq:WaveFuncR5}) and (\ref{Eq:WaveFuncR6}) we find these matrix elements. The explicit results
for both FM and AFM configurations of grains magnetizations are given in Appendix \ref{MatElApp}.

\section{Exchange interaction}
\label{Exchange}

We use perturbation theory to study the intergrain exchange interaction and search the wave function in the form
\begin{equation}\label{Eq:WaveFuncFM}
 \Psi=(1+\alpha_0)\Psi_0+\!\!\!\sum_{s,i\notin S_0^s,j\in S_0^s} \beta_{ij}^s\Psi^s_{ij}+\!\!\!\sum_{s,i\notin S_0^s,j\in S_0^s} \tilde\beta_{ij}^s\tilde\Psi^s_{ij},
\end{equation}
where $\alpha_0$, $\beta^s_{ij}$,  and $\tilde \beta^s_{ij}$ are small coefficients to be found later.
In~(\ref{Eq:WaveFuncFM}) we take into account only states with one electron transferred between the grains.
States where two electrons jumping between grains are neglected since these states
have much bigger Coulomb energy ($2\tilde \epsilon_\mathrm c$). Also, we neglect electrons
transitions between single particle states within the same grain.
These transitions do not contribute to energy within our accuracy. 
Note that we consider the grains with the radius of few nms. Such grains have 
thousands of electrons. Therefore, adding one electron can be considered as 
a perturbation. This approach is not valid for magnetic clusters consisting of few atoms.

\subsection{FM state wave function}

We start our calculations with FM state.
Using the perturbation theory and particle conservation requirement
we find the following result for coefficients
in~(\ref{Eq:WaveFuncFM}) (for details see Appendix~\ref{Formalism})
\begin{equation}\label{Eq:WaveFuncFM1}
 \beta_{ij}^s=\tilde\beta_{ij}^s=-\frac{T^s_{ij}}{\epsilon_i^s-\epsilon_j^s+\tilde \epsilon_\mathrm c}.\\
\end{equation}
\begin{equation}\label{Eq:PartNumbFM}
\alpha_0=2\!\!\!\!\!\!\!\!\!\sum_{s,i\notin S_0^s,j\in S_0^s}\!\!\frac{\mathrm{Re}(  T^s_{ij}P^{s*}_{ij})}{\epsilon_i^s-\epsilon_j^s+\tilde \epsilon_\mathrm c}-\!\!\!\!\!\sum_{s,i\notin S_0^s,j\in S_0^s}\!\!\frac{|  T^s_{ij}|^2}{(\epsilon_i^s-\epsilon_j^s+\tilde \epsilon_\mathrm c)^2}.
\end{equation}

Using the wave function $\Psi$ in (\ref{Eq:WaveFuncFM}) we can calculate the energy
of the FM state. Taking into account the fact that the mean energy level spacing is much smaller
than the Fermi energy we change summation with integration and obtain
the following result for energy (see details in Appendix~\ref{Formalism})
\begin{equation}\label{Eq:EnFM1}
\eqalign{
 E^{\mathrm{FM}}\!=\langle\Psi|\hat H|\Psi\rangle=\cr
 =\frac{\Omega}{(2\pi)^2}\sum_s\!\int^{k_\mathrm F^s}_0\!\!dk \left(\!E_\mathrm F-\!\epsilon^s_k\right)\!\left(\!\frac{E_\mathrm F+\epsilon^s_k}{2}+\frac{\pi a}{\varkappa^s}V^s_{kk}\!\right)-\cr
 -\frac{a\Omega^{2}}{8\pi^2}\sum_s\int^{k^s_\mathrm{F}}_{0}\!\!\!\!dk_1\int^{k^s_\mathrm{F}}_0\!\!dk_2 \delta^s(k_1,k_2)\frac{  \mathrm{Re}(T^s_{12}P^{s*}_{12})}{\varkappa^s_2}-\cr
 -\frac{a\Omega^{2}}{8\pi^2}\sum_s\int^{k^s_\mathrm{max}}_{0}\!\!\!\!dk_1\int^{\mathrm{min}(k_1,k_\mathrm F^s)}_0\!\!\!\!\!\!dk_2\xi^s(k_1,k_2)\times\cr
 \times \frac{|  T^s_{12}|^2/\varkappa_2^s}{\frac{\hbar^2}{2m_\mathrm e}(k_1^2-k_2^2)+\tilde \epsilon_\mathrm c}.
}
\end{equation}
Here $\epsilon^s_k=\hbar^2k^2/(2m_\mathrm e)-U+s\Ji$ and we introduce the following functions and notations
\begin{equation}\label{Eq:AuxFunc}
\xi^s(k_1,k_2)=\frac{\hbar^2}{2m_\mathrm e}\left\{\eqalign{{l}(k_1^2-k_2^2),~k_1<k_\mathrm F^s,\cr
((k_\mathrm F^s)^2-k_2^2),k_1>k_\mathrm F^s,~}\right.
\end{equation}
\begin{equation}\label{Eq:AuxNot1}
\eqalign{
 k^s_\mathrm{max}=\sqrt{2m_e(U-sJ)/\hbar^2},\cr
 k^s_\mathrm{F}=\sqrt{2m_e(E_\mathrm F+U-sJ)/\hbar^2}.
}
\end{equation}
\begin{equation}\label{Eq:AuxFunc2}
\delta^s(k_1,k_2)=\left\{\eqalign{ E_\mathrm F+U-s\Ji-\frac{\hbar^2k^2_1}{2m_\mathrm e},~k_2<k_1,\cr
E_\mathrm F+U-s\Ji-\frac{\hbar^2k^2_2}{2m_\mathrm e},~k_1>k_2.~}\right.
\end{equation}
In (\ref{Eq:EnFM1}) the matrix elements $  T^s_{12}$, $P^s_{12}$ and $V^s_{kk}$ are defined by
(\ref{Eq:ME2}) with functions $F_i$ being replaced by one.
For semimetal with only one spin subband occupied ($E_\mathrm F<\Ji-U$) we sum in (\ref{Eq:EnFM1})
only over the occupied spin subband ($s=$``-'').

\subsection{AFM state wave function}

Using the perturbation theory we find the following result for coefficients
in (\ref{Eq:WaveFuncFM})
\begin{equation}\label{Eq:WaveFuncAFM1}
\eqalign{
 \beta_{ij}^s=\tilde\beta_{ij}^{-s}=-\frac{T^s_{ij}}{\epsilon_i^{-s}-\epsilon_j^s+\tilde \epsilon_\mathrm c},\cr
 \alpha_0=2\!\!\!\!\!\!\!\!\!\sum_{s,i\notin S_0^{-s},j\in S_0^s}\!\!\frac{ \mathrm{Re}(T^s_{ij}P^{s*}_{ij})}{\epsilon_i^{-s}-\epsilon_j^s+\tilde \epsilon_\mathrm c}-\!\!\!\!\!\sum_{s,i\notin S_0^{-s},j\in S_0^s}\!\!\frac{|  T^s_{ij}|^2}{(\epsilon_i^{-s}-\epsilon_j^s+\tilde \epsilon_\mathrm c)^2}.
}
\end{equation}
The energy of the AFM state has the form
\begin{equation}\label{Eq:EnAFM1}
\eqalign{
 E^{\mathrm{AFM}}\!=\!\frac{\Omega}{(2\pi)^2}\!\sum_s\!\!\int^{k_\mathrm F^s}_0\!\!\!\!dk\!\left(\!E_\mathrm F-\epsilon^-_k\right)\!\left(\!\frac{E_\mathrm F+\epsilon^-_k}{2}+\frac{\pi a}{\varkappa^s}V^s_{kk}\!\right)\!-\cr
 -\frac{a\Omega^{2}}{8\pi^2}\sum_s\int^{k^{-s}_\mathrm F}_{0}\!\!\!\!dk_1\int^{k^s_\mathrm F}_0\!\!\!\!\!\!dk_2 \tilde \delta(k_1,k_2)\frac{\mathrm{Re}(T^s_{12}P_{12}^{s*})}{\varkappa^s_2}-
\\ -\frac{a\Omega^{2}}{8\pi^2}\left\{\int^{k^-_\mathrm{max}}_{\sqrt{2\tilde J_\mathrm{sd}}}\!\!\!\!dk_1\int^{k^-_\mathrm{up}}_0\!\!\!\!\!\!dk_2\tilde\xi^-(k_1,k_2)\right.\times\cr
 \times\frac{|  T^+_{12}|^2/\varkappa^+_2}{\frac{\hbar^2}{2m_\mathrm e}(k_1^2-k_2^2-2\tilde J_\mathrm{sd})+\tilde \epsilon_\mathrm c}+\cr
 +\!\!\!\int^{k^+_\mathrm{max}}_{0}\!\!\!\!dk_1\!\!\!\int^{k^+_\mathrm{up}}_0\!\!\!\!\!\!dk_2\tilde\xi^+(k_1,k_2)\left.\frac{|  T^-_{12}|^2/\varkappa^-_2}{\frac{\hbar^2}{2m_\mathrm e}(k_1^2-k_2^2+2\tilde J_\mathrm{sd})+\tilde \epsilon_\mathrm c}\right\}.
}
\end{equation}
Here we introduce the following functions
\begin{equation}\label{Eq:AuxFunc1}
\tilde\xi^s(k_1,k_2)=\frac{\hbar^2}{2m_\mathrm e}\left\{\eqalign{(2s\tilde J_\mathrm{sd}+k_1^2-k_2^2),~k_1<k_\mathrm F^s,\cr
((k_\mathrm F^{-s})^2-k_2^2),k_1>k_\mathrm F^s,~}\right.
\end{equation}
\begin{equation}\label{Eq:AuxFunc3}
\tilde \delta^s(k_1,k_2)=\left\{\eqalign{ E_\mathrm F+U+s\Ji-\frac{\hbar^2k^2_1}{2m_\mathrm e},~2s\tilde J_{\mathrm{sd}}+k^2_2<k^2_1,\cr
E_\mathrm F+U-s\Ji-\frac{\hbar^2k^2_2}{2m_\mathrm e},~2s\tilde J_{\mathrm{sd}}+k^2_2>k^2_1,~}\right.
\end{equation}
and
\begin{equation}\label{Eq:AuxFunc4}
k^s_\mathrm{up}=\mathrm{min}(\sqrt{k^2_1+2s\tilde J_\mathrm{sd}} \, ,k_\mathrm F^{-s}).
\end{equation}
In the above expressions we use the notation $\tilde J_\mathrm{sd}=2m_\mathrm e\Ji/\hbar^2$.

The semimetal case should be divided into two limits: 1) $\Ji<U$; and 2) $\Ji>U$. In the first case we need to consider
only one spin subband ($s=$``-'') in (\ref{Eq:EnAFM1}). In the second case, all hopping terms (the second and the third
terms in (\ref{Eq:EnAFM1})) are zero. As a result we find for $J_\mathrm{sd}>U$
\begin{equation}\label{Eq:EnAFMssb1}
 E^{\mathrm{AFM}} = \frac{\Omega}{(2\pi)^2}\int^{k_\mathrm F^-}_0dk \left(\!E_\mathrm F-\epsilon^-_k\right)\!\left(\!\frac{E_\mathrm F+\epsilon^-_k}{2}+\frac{\pi a}{\varkappa^s}V^s_{kk}\!\right).
\end{equation}
We can calculate the intergrain exchange interaction as
difference between $E^\mathrm{AFM}$ and $E^\mathrm{FM}$ using (\ref{Eq:ExFin}).

\section{Discussion of results}\label{Sec:Anal}

\subsection{Granular magnets}

Granular magnet is an ensemble of magnetic grains.
The exchange interaction between the grains leads to the formation of long-range magnetic
order. While we calculate the intergrain exchange interaction at
zero temperature, our results for $J$ are valid for temperatures below the charging energy,
$T\ll \tilde\epsilon_\mathrm c$ where temperature fluctuations of $J$ can be neglected.
However, temperature fluctuations can not be neglected in discussing
the magnetic long-range order in granular magnets since the temperature
can be comparable with exchange coupling, $T\approx J\ll\tilde\epsilon_\mathrm c$.
Magnetic structure of GFM is defined by the ratio $T/J$. Thermal fluctuations
destroy the long range magnetic order above a certain
temperature which is called the ordering temperature,
$T_\mathrm{ord}$.~\cite{Glazman2002, Freitas2001, Sobolev2012, Hembree1996}

Beside temperature fluctuations and the intergrain exchange interaction
the magnetic state of granular magnets is defined by magnetic anisotropy of
a single grain~\cite{Chien1988, Chien1991} and by the intergrain
magneto-dipole (MD) interaction~\cite{Kechrakos98, Grady1998, Ayton95, Ravichandran96, Mamiya99, Djurberg97, Sahoo03}.
The magnetic anisotropy leads to blocking phenomena while MD interaction due to its
long-range nature forms the spin glass state. Below we neglect the MD interaction
assuming that the grain size is small enough. The influence of MD interaction on the magnetic state of GFM was
discussed in Refs.~\cite{Ayton95, Ravichandran96, Mamiya99, Djurberg97, Sahoo03}.

The exchange interaction can lead to different types of macroscopic magnetic states depending
on its sign. For FM interaction, $J > 0$, the long-range magnetic order is
the SFM state with finite magnetization and coercive field. For AFM interaction, $J < 0$, in
the presence of disorder the spin glass state is realized.

The SFM state can be studied using the mean field approach
where all magnetic grains have a strong uniaxial anisotropy leading to only
two magnetic states for each grain (Ising model)~\cite{Sobolev2012, Munakata2009}.
All anisotropy axes have the same direction. The exchange interaction $J$ between
the grains is finite and the MD interaction is zero.
In this model the ordering temperature and the intergrain exchange interaction are related as follows
$T_{\mathrm{ord}}=z_\mathrm n J$, where $z_\mathrm n$ is the coordination number,
$z_\mathrm n=6$ for three dimensional cubic lattice.
Below we will discuss the intergrain exchange interaction based on this model.
All figures will show the quantity $z_\mathrm n J$ which is related to the
measurable parameter $T_\mathrm{ord}$.

\subsection{Influence of the Coulomb interaction}

Equations~(\ref{Eq:EnFM1}) and (\ref{Eq:EnAFM1}) show that there are three different
contributions to the intergrain exchange interaction. All these contributions exist in
Slonczewski and Bruno models for exchange interaction
between FM layers separated by an insulating spacer \cite{Bruno1995,Slonczewski1989}.
The first two contributions do not depend on the charging energy, $\tilde\epsilon_\mathrm c$.
As a result these terms are not affected by the Coulomb interaction. The third contribution is due to virtual electron
hopping between the grains: the hop of electron from one grain into another results in charging of both grains.
Therefore the hopping contribution involves virtual states with charged grains.
The energy of these virtual states has an additional contribution due to the
presence of operator $\hat H_\mathrm C$ ($\tilde \epsilon_\mathrm c$).
Transitions into these virtual states are suppressed for large charging energies $\tilde \epsilon_\mathrm c$
and allowed for small energies. Varying the charging
energy $\tilde\epsilon_\mathrm c$ one can control the intergrain exchange interaction.
The charging energy, $\tilde \epsilon_\mathrm c$ depends on the grain size and matrix dielectric constant.
This effect is absent in Slonczewski model since the size of magnetic leads in this
model is infinite leading to zero charging energy ($\tilde\epsilon_\mathrm c=0$)
and disappearance of the Coulomb interaction term in the Hamiltonian. Therefore only
the height and thickness of the barrier define the exchange interaction in Slonczewski model.

The finite grain size $a$ influences the exchange interaction in two ways:
1) through the contact area between grains and 2)  through the Coulomb interaction.
For spherical grains the contact area is $S_\mathrm c\approx \pi a/\varkappa_0$.
Therefore the final result for the exchange interaction has a factor $a$.
For zero charging energy, $\tilde\epsilon_\mathrm c=0$ the exchange interaction, $J$
depends linearly on $a$ in contrast to bulk FM separated by the insulating layer, where $J\sim a^2$. The third term in the exchange interaction in (\ref{Eq:EnFM1}) and (\ref{Eq:EnAFM1}) results in positive (FM) contribution to the exchange coupling. Therefore, decreasing $a$ (enhancing the Coulomb blockade effect) one can decrease the FM contribution to the intergrain
exchange coupling and shift the coupling toward the AFM type. One can even observe the
transition between the FM and AFM exchange coupling changing the charging energy, $\tilde\epsilon_\mathrm c$.

To demonstrate the dependence of the intergrain exchange
interaction on the grain size and the dielectric constant we use the approximate analytical formula
in (\ref{Eq:ExFin}) instead of complicated integrals. We find the following
approximate expression for the intergrain exchange interaction
\begin{equation}\label{Eq:ExHopEps}
J=S_\mathrm c J_{0}+S_\mathrm c J_{1}\left(\!\!1-\sqrt{\frac{\tilde\epsilon_{\mathrm c}}{\Delta\epsilon}}\mathrm{arctan}\left(\sqrt{\frac{\Delta\epsilon}{ \tilde\epsilon_{\mathrm c}}}\right)\!\!\right),
\end{equation}
where $\Delta\epsilon=\gamma\hbar\sqrt{h_\mathrm B}/(\sqrt{2m}d)$ is the
characteristic energy interval (around the Fermi level) contributing to the
hopping based exchange interaction (see Ref.~\cite{Slonczewski1989}),
$\gamma\approx 3.43$, the barrier height $h_\mathrm B=-E_\mathrm F$.
Parameters $J_0$ and $J_1$ ($J_1>0$) can be
considered as the areal exchange interaction. Parameters $J_{0,1}$ depend on $E_\mathrm F$, $U$ and $J_{sd}$,
but do not depend on the dielectric permittivity $\varepsilon$ and the grain size $a$.
The grain size enters in (\ref{Eq:ExHopEps}) in the contact area $S_\mathrm{c}\sim a$
and the charging energy $\tilde\epsilon_\mathrm c\sim 1/a$.
The dielectric permittivity in this equation also enters through
the quantity $\tilde\epsilon_\mathrm c \sim 1/\varepsilon$.

Equation~(\ref{Eq:ExHopEps}) shows that the Coulomb interaction
becomes important only when the charging energy becomes comparable or
larger than the interval $\Delta\epsilon$. Thus, to investigate
the Coulomb blockade effects it is better to use a thick insulator layer
with low barrier, instead of thin insulator with high barrier.
Therefore, in the next subsections we will consider the case of low barrier.

\subsection{Exchange interaction vs matrix dielectric constant}

The part of the exchange coupling depending on $\varepsilon$ can be described as $\sqrt{\tilde\epsilon_{\mathrm c}/\Delta\epsilon} \, \mathrm{arctan}(\sqrt{\Delta\epsilon/\tilde\epsilon_{\mathrm c}})$. This function is zero for infinite $\varepsilon$ and tends to 1 as $\varepsilon\to 0$. Obviously, $\varepsilon>1$. Thus, the exchange
interaction grows with increasing the dielectric constant $\varepsilon$.
Below we compare the intergrain exchange interaction calculated
for finite dielectric constant ($\epsilon \approx 5$)
and for $\epsilon\to \infty$ (the limit of zero charging energy)
\begin{equation}\label{Eq:ExHopEps1}
\Delta J=J|_{\varepsilon=5}-J|_{\varepsilon=\infty}=-S_\mathrm c J_{1}\sqrt{\frac{\tilde\epsilon_{\mathrm c}}{\Delta\epsilon}}\mathrm{arctan}\left(\sqrt{\frac{\Delta\epsilon}{ \tilde\epsilon_{\mathrm c}}}\right)\!\!.
\end{equation}

For $J_0\gg J_1$ one can write $\Delta J/J\approx (J_1/J_0)\sqrt{\tilde\epsilon_{\mathrm c}/\Delta\epsilon}\, \mathrm{arctan}(\sqrt{\Delta\epsilon/\tilde\epsilon_{\mathrm c}})$.
The characteristic region of energies contributing to the exchange
interaction in (\ref{Eq:EnFM1}) and (\ref{Eq:EnAFM1}) is
$\Delta \epsilon\approx 120$ K for $E_\mathrm F=-0.1$ eV and $d\approx 1$ nm. The
charging energy is  $\tilde\epsilon_\mathrm c\varepsilon a=0.8$ eV$\cdot$nm. The
charging energy for grains with $a=2.5$ nm and $\varepsilon=5$ is about 800 K resulting in
$\sqrt{\tilde\epsilon_{\mathrm c}/\Delta\epsilon}\, \mathrm{arctan}(\sqrt{\Delta\epsilon/\tilde\epsilon_{\mathrm c}})\approx 0.95$.
In this case one has $\Delta J/J\approx J_1/J_0$. The maximum variation of
the exchange interaction is defined by constants $J_0$ and $J_1$.
\begin{figure}
\includegraphics[width=0.5\columnwidth]{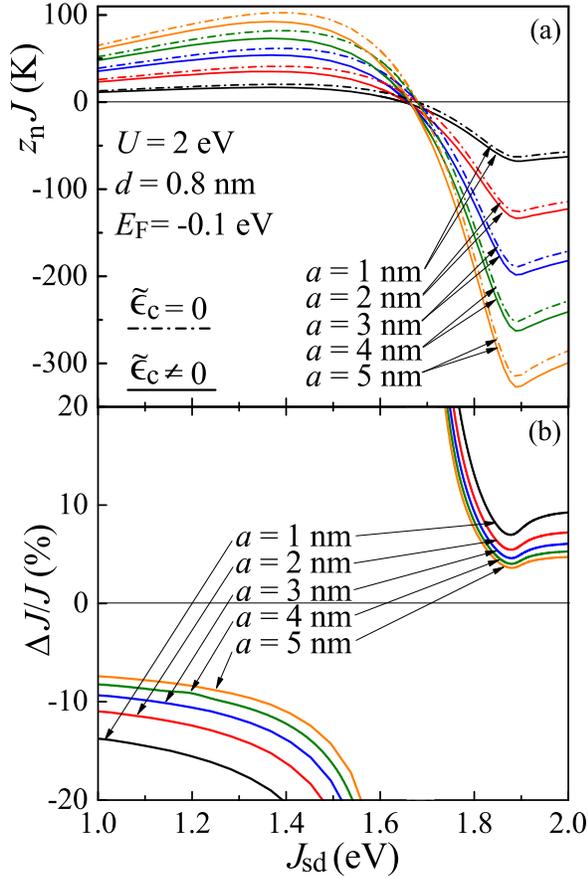}
\caption{(Color online) (Upper panel) Exchange interaction vs grain size $a$ and spin subband
splitting $\Ji\in[1;2]$ eV for $d=0.8$ nm, $U=2$ eV, and $E_\mathrm F=-0.1$ eV and coordination number
$z_\mathrm n=6$. (Lower panel) The relative difference of the intergrain exchange
coupling $\Delta J/J$
vs spin subband splitting $\Ji$ for different grain sizes $a$.} \label{Fig:JvsJ1}
\end{figure}

Figure~\ref{Fig:JvsJ1} shows a variation of the exchange interaction
between magnetic grains embedded into insulating matrix for different dielectric
constants and the same barrier height. The curves are
calculated using (\ref{Eq:EnFM1}) and (\ref{Eq:EnAFM1}).
The upper panel shows the intergrain exchange interaction vs spin subband splitting $\Ji$
for the following parameters:
barrier height $h_\mathrm B=-E_\mathrm F=0.1$ eV, $U=2$ eV, $d=0.8$ nm.
Several curves are shown for different grain sizes ranging from 1 to 5 nm.
Solid lines correspond to finite charging energy,
$\tilde \epsilon_\mathrm c=0.14/a$ eV ($a$ is in nm) and $\varepsilon= 5$.
Dash-dotted curves correspond to infinite dielectric constant where the
Coulomb blockade is negligible, $\tilde \epsilon_\mathrm c\approx0$.
For any grain sizes the exchange interaction has a peak in the vicinity of $\Ji=0.7 U$.
The peak value grows with the grain size $a$. The exchange coupling is strong and exceeds 100 K
for grains with radius 5 nm. Thus, transition between the SPM and SFM states in granular magnets
is experimentally observable. For
$J_{\mathrm{sd}}>(E_\mathrm F+U)$ the exchange interaction becomes of AFM type.
The absolute value of AFM coupling reaches its maximum at $J_\mathrm{sd}\approx 0.9 U$.

One can see that the change in the exchange interaction with changing
the insulating dielectric constant is pronounced for grains with $a=5$ nm.
The ordering temperature variation due to the Coulomb blockade is of
the order of 10-20 K. The curves with zero charging
energy $\tilde\epsilon_\mathrm c$ are located above the curves with
finite $\tilde\epsilon_\mathrm c$ meaning that the third ``hopping'' term in the exchange interaction
in (\ref{Eq:EnFM1}) and (\ref{Eq:EnAFM1}) results in positive FM contribution.
Therefore, the Coulomb blockade effects, which are pronounced for small dielectric
constants $\varepsilon$ and small grains, reduce the FM coupling between the grains.

The lower panel shows the relative difference (in \%) between
the solid and the dash-dotted lines of the upper panel,
$\Delta J/J$, calculated using (\ref{Eq:EnFM1}) and (\ref{Eq:EnAFM1}).
These curves show that the relative change of the intergrain exchange interaction due to the Coulomb
interaction can be a few tens of percent. The relative change grows with decreasing the grain size meaning
that the Coulomb blockade effects are more pronounced for small grains.

\begin{figure}
\includegraphics[width=0.5\columnwidth]{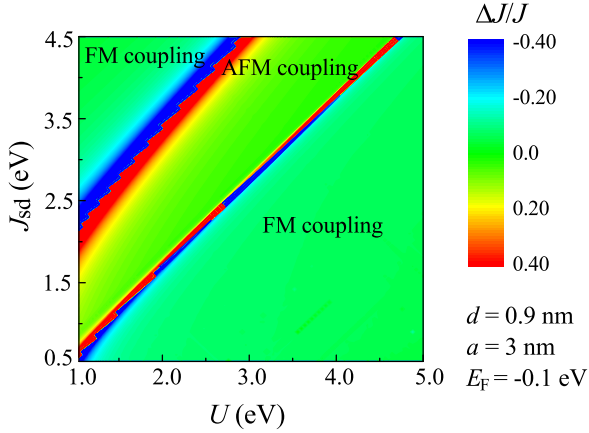}
\caption{(Color online) Relative change of exchange
interaction
$\Delta J/J$
vs $U$
and spin subband splitting $\Ji$ for $d=0.9$ nm, $a=3$ nm, and $E_\mathrm F=-0.1$ eV.} \label{Fig:Peak}
\end{figure}

Figure~\ref{Fig:Peak} shows the relative change of the exchange
interaction $\Delta J/J$
vs $U$ (or the Fermi momentum $k^2_\mathrm F=2m_\mathrm e(U+E_\mathrm F)/\hbar^2$) and the
spin subband splitting $\Ji$.
Two bright (blue and red) diagonal lines appear in the vicinity of the
zero exchange. These lines divide the whole space into regions of AFM and FM coupling.
The relative change of the exchange interaction $\Delta J/J$ grows sharply,
reaching infinity when $J|_{\tilde \epsilon_\mathrm c\ne0}+J|_{\tilde \epsilon_\mathrm c=0}=0$ ($J_0\approx 0$).
This produces the red and the blue spots along the two diagonals in figure~\ref{Fig:Peak}.

In general, the relative difference grows with reducing the
Fermi momentum (or $U$). This can be understood as follows:
The exchange interaction is the result of virtual electron hopping between the grains.
The hopping is related to the kinetic energy of electrons with characteristic
energy scale $E_\mathrm F$. The larger the ratio $E_\mathrm C/E_\mathrm F$ the
stronger the influence of many-body effects on hopping and thus on
the intergrain exchange coupling. Since the exchange interaction is
stronger for large spin subband splitting ($J_\mathrm{sd}>(U+E_\mathrm F)$)
the influence of the Coulomb interaction is more pronounced in this region too.

To observe the influence of the Coulomb interaction
on the exchange interaction $J$ experimentally one can either use the
insulating matrix with different dielectric constants or use materials
with dielectric constant being dependent on some parameter.
The first approach is difficult since insulators with different
dielectric constants $\varepsilon$ have different electron energy barriers.
The second approach looks more promising. For example, one can use
ferroelectrics with temperature or field dependent dielectric constant to control the
magnetic interaction in GFM.

\subsection{Exchange interaction vs grain size $a$ and the spin subband splitting $\Ji$}

\begin{figure}
\includegraphics[width=0.5\columnwidth]{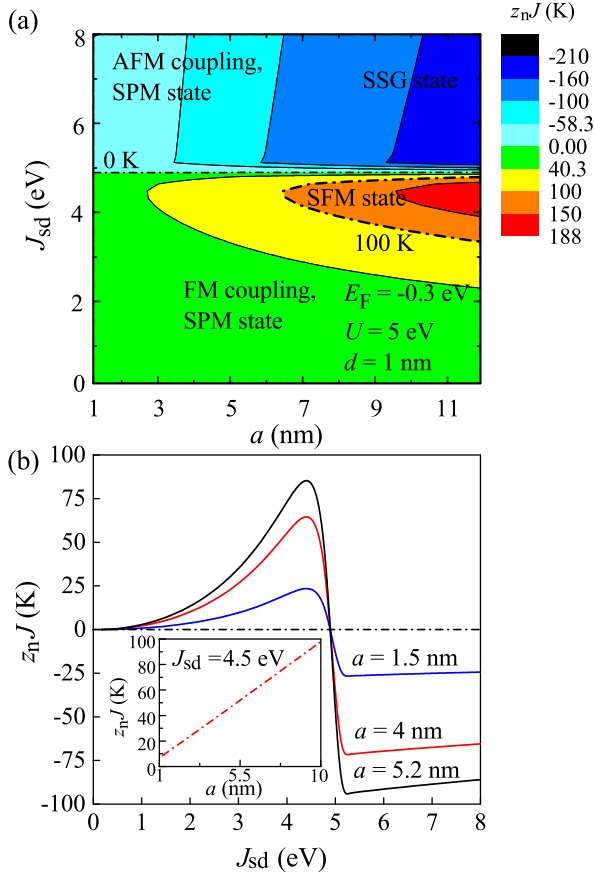}
\caption{(Color online) (a) Magnetic phase diagram of granular magnet
as  a function of grain size $a$ and the spin subband splitting $\Ji$ for the following
parameters: $d=1$ nm, $U=5$ eV, and $E_\mathrm F=-0.1$ eV. Color shows the value
of the product of intergrain coupling $J$ and coordination number $z_\mathrm n=6$. The region of
SFM state ($z_\mathrm n J>T$) is shown for temperature $T=100$ K. For negative $J$ the SSG state
appears at some temperature. The SPM state is shown for $|z_\mathrm n J|<T$. (b) Intergrain exchange
coupling $J$ vs spin subband splitting $\Ji$ for different grain sizes $a$. Inset: intergrain
exchange interaction $J$ vs the grain size $a$ for  $\Ji=4.5$ eV.} \label{Fig:Diagram1}
\end{figure}

The Coulomb blockade effect depends on the grain size $a$:
the smaller the grain size the stronger the Coulomb blockade.
Figure~\ref{Fig:Diagram1} shows the intergrain exchange interaction vs grain size $a$ and
the spin subband splitting, $\Ji$ for given intergrain distance $d=1$ nm, barrier
height $h_\mathrm B=0.1$ eV and $U=5$ eV.
We assume that the effective dielectric constant of the medium outside the grains is $\varepsilon=5$.
This value corresponds to $a\tilde\epsilon_\mathrm c=0.14$ eV$\cdot$nm.
Figure~\ref{Fig:Diagram1} shows that the coupling between the grains can be either FM or AFM.
The dash-dotted line ($J=0$ K) in figure~\ref{Fig:Diagram1} divides the regions of FM and AFM coupling:
the small spin subband splitting $\Ji$ results in the FM intergrain coupling, while the strong
splitting leads to the AFM exchange interaction between the grains. Transitions between the
two regions occur at
$J\approx (U+E_\mathrm F)$.
\begin{figure}
\includegraphics[width=0.5\columnwidth]{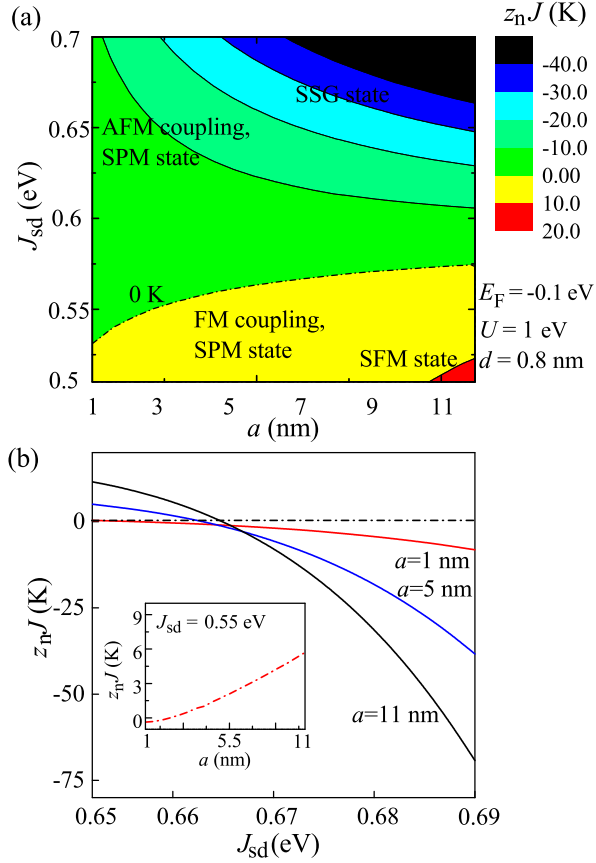}
\caption{(Color online) (a) Magnetic phase diagram of granular magnet as
a function of grain size $a$ and the spin subband splitting $\Ji$ for
$d=0.8$ nm, $U=1$ eV, and $E_\mathrm F=-0.1$ eV. Color shows the value of
$z_\mathrm nJ$. The regions of SFM ordering
($z_\mathrm n J>T$) is shown for temperature $T=20$ K. For large enough
negative value of $J$ the super-spin glass state may appear.
The SPM state appears at $|z_\mathrm n J|<T$. (b) Intergrain exchange
coupling vs spin subband splitting $\Ji$ for different grain sizes $a$. Inset:
intergrain exchange coupling $J$ vs the grain size for $\Ji=0.55$~eV.} \label{Fig:Diagram2}
\end{figure}

For large values of Fermi momentum $k_\mathrm F$  the
transition from FM to AFM coupling does not depend on the grain size ($J_1\ll J_0$). Therefore the
$J=0$ K line is straight and parallel to the horizontal axis. The
kinetic energy of electrons in this case exceeds the Coulomb energy
reducing the role of the Coulomb blockade effects.

The long-range SFM order in granular array appears when the product
$z_\mathrm n J$ reaches the system temperature $T$. This region is shown for temperature $T=100$ K.
 A strong AFM coupling leads to the formation of SSG state in
disordered granular magnets. Inset in figure~\ref{Fig:Diagram1}(b) shows the
intergrain coupling vs the grain size.
For high electron concentration, large $k_\mathrm F$, the dependence is linear. 
In this case the influence of the third ``hopping'' term in (\ref{Eq:EnFM1})
and (\ref{Eq:EnAFM1}) on the intergrain exchange interaction is small
and the Coulomb blockade does not influence the exchange interaction.
This result corresponds to the case of grains made of strong FM metals
such as Fe, Ni or Co with wide conduction band.

Note that the exchange coupling $J$ is the total interaction 
energy between two grains. It linearly grows with the grain size. At the same time the
areal interaction energy (exchange coupling per unit surface area, or even per atom) 
does not grow with the grain size. Moreover, the areal exchange coupling even 
decreases with $a$ since the effective interaction area ($\sim a\varkappa$) is much smaller 
than the total grain surface ($\sim a^2$). This is due to spherical shape of the grains.

\begin{figure}
\includegraphics[width=0.5\columnwidth]{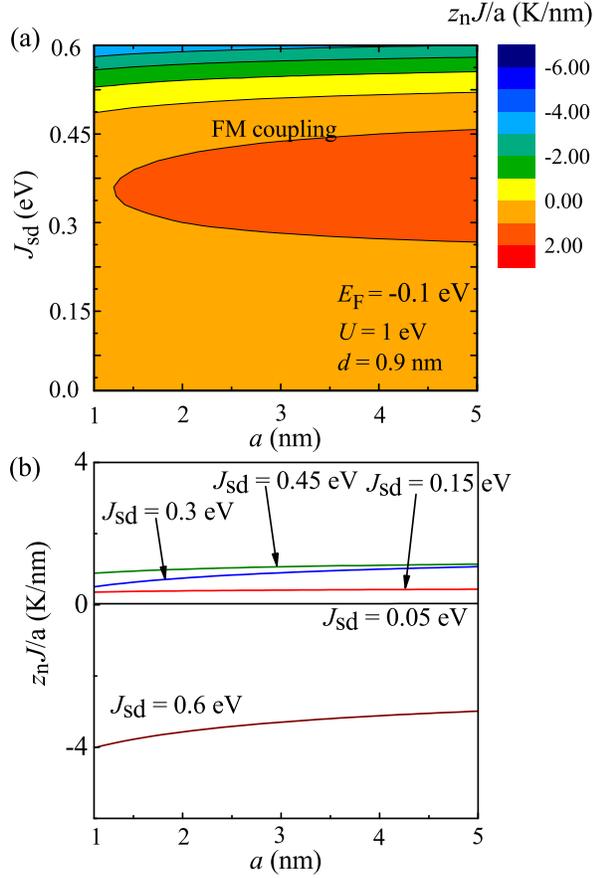}
\caption{(Color online) (a) Magnetic phase diagram: normalized exchange
interaction $z_\mathrm n J/a$ vs the grain size $a$ and the spin subband
splitting $\Ji$ for $d=0.9$ nm, $U=1$ eV, and $E_\mathrm F=-0.1$ eV. Color
shows the value of $z_\mathrm n J/a$. (b) Normalized exchange
interaction $z_\mathrm n J/a$ vs grain size $a$ for different spin subband
splitting $\Ji$.} \label{Fig:DiagramNorm}
\end{figure}

Figure~\ref{Fig:Diagram2} shows the intergrain exchange coupling vs grain size $a$ and the spin subband splitting $\Ji$
for a given intergrain distance $d=0.8$ nm, barrier height $h_\mathrm B=0.1$~eV,
$U=1$ eV, and  $a\tilde \epsilon_\mathrm c=0.14$ eV$\cdot$nm
(small Fermi momentum and low electron concentration).
For small Fermi momentum the Coulomb interaction can substantially
modify the intergrain exchange coupling. In this case the curve $J=0$ K is no longer a straight line.
Figure~\ref{Fig:Diagram2} shows that the exchange coupling changes its sign from FM to AFM with reducing the grain size
at fixed $\Ji$ (see inset in figure~\ref{Fig:Diagram2}(b)).

To observe the influence of the Coulomb interaction on the exchange coupling experimentally
one can measure the normalized SFM ordering temperature as a function of grain size $a$.
In the absence of Coulomb blockade the exchange interaction $J$ is a linear function of $a$
and so is the ordering temperature, $T_\mathrm{ord}=z_\mathrm n J\sim a$.
In the presence of Coulomb blockade the dependence $J(a)$ is more complicated.
The ratio $T_\mathrm{ord}/a$ as a function of $a$ provides information about the
influence of the Coulomb blockade. For zero Coulomb interaction the normalized ordering temperature
is constant and does not depend on $a$. The Coulomb interaction results in deviation
of $T_\mathrm{ord}/a$ from the straight line. Figure~\ref{Fig:DiagramNorm} shows the
normalized exchange interaction (or ordering temperature) as a function of grain
size $a$ and spin subband splitting $\Ji$. Figure~\ref{Fig:DiagramNorm}(b) shows the
normalized ordering temperature as a function of $a$ for different $\Ji$.

The influence of Coulomb blockade on the intergrain exchange
interaction is especially pronounced in GFM made of FM metals with small
Fermi momentum and large spin subband splitting. Halfmetals
with low electron concentration at the Fermi level and full spin
subband splitting such as CrO$_2$ and Sr$_2$FeMoO$6$~\cite{Schwarzt1986,Tokura1998}
would be good candidates for observation of many-body effects in GFM.

\section{Validity of our model}\label{Sec:Val}

Below we discuss several assumptions and approximations of our theory.

1) In our consideration we took into account only elastic transitions of electrons and neglected
the inelastic sequential tunneling between grains. At finite temperature sequential
tunneling also contribute to the exchange interaction. Elastic transitions contribute to the exchange
coupling only in the second order perturbation theory (in the tunneling matrix element), while
sequential tunneling contributes in the first order perturbation theory.
In our consideration sequential tunneling can be neglected since these
transitions are exponentially suppressed due to the Coulomb blockade effects.
The electron spectrum in the grains has a Coulomb gap $\tilde\epsilon_\mathrm c$ leading
to the exponential suppression of inelastic sequential tunneling, $\rme^{-\tilde\epsilon_\mathrm c/T}$.
Elastic processes are also suppressed due to the gap but only algebraically,
$1/\tilde\epsilon_\mathrm c$. Similar situation exists in granular metals~\cite{Bel2007review}:
Elastic co-tunneling, appearing in the second or higher order perturbation theory, exceeds at low
temperatures the inelastic sequential tunneling appearing in the first order perturbation theory.

2) We discussed the influence of the diagonal spin-independent
part of the Coulomb interaction on the intergrain exchange coupling
and neglected the spin-dependent part of the Coulomb
interaction between electrons located at different grains. However, the
overlap of electron wave functions in the insulator between the grains produces
finite spin dependent matrix elements of the Coulomb interaction, which originally
was called the exchange coupling~\cite{Landau3}
\begin{equation}\label{Eq:ExMatEl}
U^{\mathrm{ex}}_{ijkl}=\!\int\!\!\int d^3\rv_1 d^3 \rv_2 \psi^*_i(\rv_1)\phi_l(\rv_1)\hat U_\mathrm C \psi_j(\rv_2)\phi^*_k(\rv_2),
\end{equation}
where $\hat U_\mathrm C$ is the Coulomb interaction operator. This term requires a separate consideration.

3) We assumed the parabolic electron spectrum in our model.
In our theory the hopping based exchange interaction can be either AFM or
FM depending on the shift of the spin subband.
Taking into account the real band structure via \textit{ab initio} modeling
will provide an additional insight into the problem,
however it requires more complicated calculations.
The \textit{ab initio} calculations of band structure of FM metals
can be used to estimate the effective parameters in the Hamiltonian in
(\ref{Eq:HamIn}). Also these parameters can be estimated using
spectroscopic experiments and experiments on tunneling magneto-resistance
in magnetic tunnel junctions.

4) We note that the theory developed in this paper fails as the intergrain distance 
tends to zero. Decrease in intergrain distance leads to the increase in tunneling probability. 
At some point the approximation of wave functions localized at different grains is not valid. 
In this case the better starting point is to use the wave functions delocalized 
on the scale of the whole granular magnet.

\section{Conclusion}

We developed a theory of the exchange interaction between FM metallic grains embedded into insulating matrix
by taking into account many-body effects. In particular, we considered the Coulomb blockade effects.
These effects can be neglected for layered structures, however they are crucial for nanogranular systems.
For bulk ferromagnets separated by the insulating layer the exchange interaction depends on the
barrier height for electrons inside the insulator.
We showed that due to the Coulomb blockade effects the exchange coupling between FM grains embedded into
insulating matrix additionally depends on the dielectric properties of this matrix.
In particular, the FM coupling decreases with decreasing the dielectric permittivity of insulator.
The Coulomb blockade effects prevent virtual transitions of electrons between the grains
and shift the intergrain coupling toward the AFM type.
We showed that the variation in the exchange interaction due to
the Coulomb blockade effects can be a few tens of percent.

We studied the behavior of intergrain exchange interaction $J$ as a function of grain size $a$.
We showed that there are two factors defining this behavior:
1) The geometrical factor - the increase in the grain size leads to the increase in
the interaction strength. We find that in contrast to layered structure, where exchange
interaction grows linearly with the surface area of the system, $J\sim a^2$, in granular system the exchange coupling
depends linearly on the grain size, $J\sim a$.
2) The influence of grain size on the Coulomb blockade effects and thus
on the intergrain exchange interaction. The smaller the grain the larger the Coulomb blockade
the smaller the FM contribution to the exchange interaction.
We found that the exchange interaction decays faster than the first power of $a$ with decreasing the grain size
and showed that the transition from the FM to AFM coupling exists with decreasing the grain size.

The Coulomb blockade effects are important if charging energy essentially
exceeds the characteristic energy interval contributing to the exchange interaction.
This interval depends on the barrier height of the insulator
and barrier thickness. The influence of Coulomb blockade effects is more
pronounced for thick and low barrier.
We investigated the intergrain exchange coupling as
a function of system parameters such as internal
subband splitting and the Fermi energy. The Coulomb blockade influences
the intergrain exchange coupling strongly if: 1) the Fermi momentum is not too large
and 2) the spin subband splitting is of the order of the Fermi energy.

\section{Acknowledgements}

This research was supported by NSF under Cooperative Agreement Award EEC-1160504,
the U.S. Civilian Research and Development Foundation (CRDF Global)  and NSF PREM Award.
O.U. was supported by Russian Science Foundation (Grant  16-12-10340).

\appendix

\section{Single grain wave functions}
\label{grainfunctionsApp}

Consider single spherical metallic grain with radius $a$. The
orbital part of the electron wave function in a state with orbital quantum numbers ($mln$) and a spin state $s$
has the form
\begin{equation}\label{Eq:WaveFuncSpher}
\psi^s_{mln}=R^s_{nl}(r)Y_{ml}(\varphi,\theta),
\end{equation}
where $Y_{ml}(\varphi,\theta)$ is the spherical function and $R^s_{nl}(r)$ is governed by the equation
\begin{equation}\label{Eq:WaveFuncR}
\frac{1}{r^2}\frac{\partial}{\partial r}\left(r^2\frac{\partial R^s_{nl}}{\partial r}\right)-\frac{l(l+1)R^s_{nl}}{r^2}+
\frac{2m_\mathrm e}{\hbar^2}(E^s_n-\hat U_{\mathrm{sp}})R^s_{nl}=0.
\end{equation}
Here $U_\mathrm{sp}$ stands for $s$ component of either $\hat U_1+\hat H_{1\mathrm{m}}$ or
$\hat U_2+\hat H_{2\mathrm{m}}$ depending on the grain.
The largest contribution to the  intergrain exchange interaction appears due to electrons in the vicinity of the Fermi surface. Using the quasiclassical
approximation~\cite{Landau3} for these electrons outside the metallic grain
we find
\begin{equation}\label{Eq:WaveFuncR1}
\eqalign{
 R^s_{nl}(r)=\frac{C\hbar}{r\left(2m_\mathrm e\left(\frac{\hbar^2l(l+1)}{2m_\mathrm e r^2}-E^s_n \right)\right)^{1/4}}\times \cr \times\exp\left(-\frac{1}{\hbar}\int_a^r \sqrt{2m_\mathrm e\left(\frac{\hbar^2l(l+1)}{2m_\mathrm er^2}-E^s_n\right)}dr\right),
}
\end{equation}
where $E^s_n$ stands for the total energy of electron in the radial state $n$ and the spin state $s$.
We assume that the grain size $a$ is larger than the distance between the grains surfaces $d$. Therefore we will neglect the dependence of the effective potential on $r$ in between the grains
\begin{equation}\label{Eq:WaveFuncR2}
\eqalign{
 R^s_{nl}(r)=\frac{C\hbar}{r\left(2m_\mathrm e\left(\frac{\hbar^2l(l+1)}{2m_\mathrm e a^2}-E^s_n\right)\right)^{1/4}}\times \cr  \times\exp\left(-\frac{1}{\hbar}(r-a)\sqrt{2m_\mathrm e\left(\frac{\hbar^2l(l+1)}{2m_\mathrm ea^2}-E^s_n\right)}\right).
}
\end{equation}
For the left grain wave function $\psi$ we introduce $r=\sqrt{r_\perp^2+(a+d/2+z)^2}$ and
for the right grain wave function $\phi$ we introduce $r'=\sqrt{r_\perp^2+(a+d/2-z)^2}$.
We use the notation $\varkappa^s_{nl}=-\frac{1}{\hbar}\sqrt{2m_\mathrm e\left(\frac{\hbar^2l(l+1)}{2m_\mathrm ea^2}-E^s_n\right)}$.
For small $r_\perp$ we find
\begin{equation}\label{Eq:WaveFuncR3}
R^s_{nl}(z,r_\perp)\approx\frac{C}{a\sqrt{\varkappa^s_{nl}}}\exp\left(-\varkappa^s_{nl}\left(\frac{d}{2}\pm z+\frac{r_\perp^2}{2a}\right)\right),
\end{equation}
where the sign $\pm$ corresponds to the wave function of the left and right grains and $C$ is the normalization constant.

Inside the grain the wave function consists of two waves propagating outward and toward the particle centre.
\begin{equation}\label{Eq:WaveFuncR4}
R^s_{nl}(z,r_\perp\!)\!\approx\!\!\frac{C_\mathrm i}{a\sqrt{k^s_{nl}}}\!\!\left(\rme^{\rmi k_{nl}^s\left(\!\!\frac{d}{2}\pm z+\frac{r_\perp^2}{2a}\right)}\!\!\!+\xi^s_{nl}
\rme^{-\rmi k^s_{nl}\left(\!\!\frac{d}{2}\pm z+\frac{r_\perp^2}{2a}\right)}\!\!\right)\!\!.
\end{equation}
Here $\xi^s_{nl} = \frac{k_{nl}^s-\rmi \varkappa^s_{nl}}{k_{nl}^s+\rmi \varkappa^s_{nl}}$  is the amplitude of the reflected electron wave,
$k^s_{nl}=-\frac{1}{\hbar}\sqrt{2m_\mathrm e\left(U-s\Ji+\frac{\hbar^2l(l+1)}{2m_\mathrm ea^2}-E^s_n\right)}$, and
$C_\mathrm i$ is the normalization constant.
Below we will use the symbols $i$ and $j$ to describe a set of quantum numbers characterizing the
orbital motion of electrons. The overlap of wave functions of electrons $i$ and $j$ located in different grains exists only between the grains in a small region in the vicinity of $r_\perp=0$.
The in-plane area (($x,y$)-plane) of the overlap region is $S^{ij}_\mathrm c=\pi (\lambda^{ij}_\perp)^2$, where $\lambda^{ij}_\perp=\sqrt{2a/(\varkappa_i+\varkappa_j)}$. For electrons at the Fermi level we introduce the size $\lambda_\perp=\sqrt{a/\varkappa_0}$
and the corresponding area $S_\mathrm c=\pi\lambda^2_\perp$. \
This estimate for overlap region works for $d \ll a$ and $S_\mathrm c \ll \pi a^2$.

\begin{figure}
	\includegraphics[width=0.5\columnwidth]{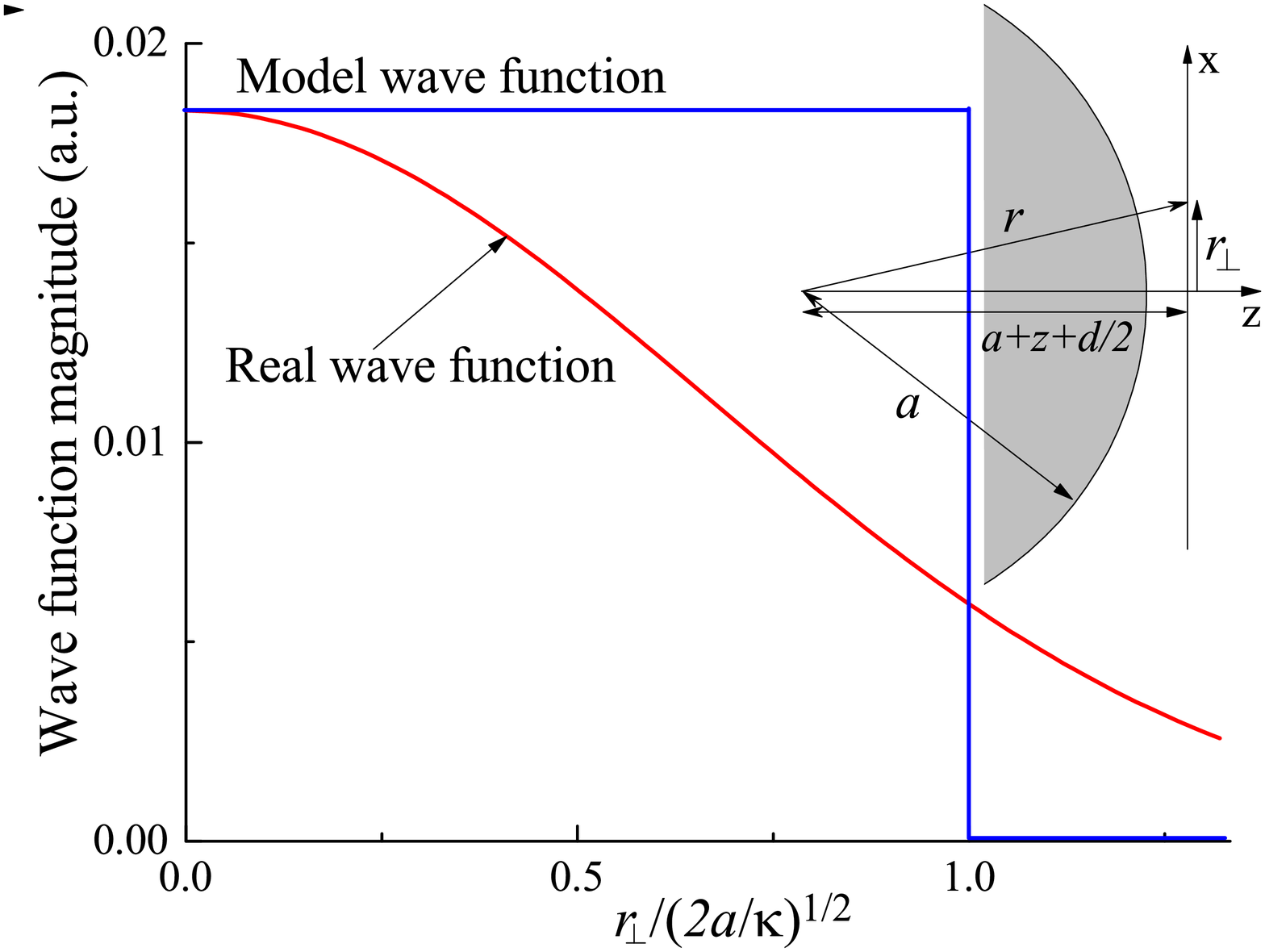}
	\caption{(Color online) Wave function magnitude vs $r_\perp$. Red line shows the wave function given by (\ref{Eq:WaveFuncR3}). Blue line shows the model wave function used in the calculation of the matrix elements.} \label{Fig:WF1}
\end{figure}

 To further simplify the wave function we change the 
 spherical wave with the plane one. The wave function magnitude in (\ref{Eq:WaveFuncR3}) exponentially decays in the ($x,y$) plane. We change it with the  wave function having constant magnitude in the region $r_\perp<\sqrt{2a/\varkappa}$. 
 This is schematically shown in figure~\ref{Fig:WF1}. Instead of factor $\rme^{-\varkappa r_\perp^2/(2a)}$ we put 1 in the region $r_\perp<\sqrt{2a/\varkappa}$ and 0 outside the region. Generally, one can use numerical calculations to avoid this simplification.

The important difference with the case of infinite semispaces
is related to the fact that the interaction between electrons at
different grains occurs only in the small area $S^{ij}_\mathrm c$. This area is small in comparison with
the grain surface and grows linearly with the grain size $a$ (instead of $a^2$). This geometric factor
decreases the interaction between the grains.

The size of the interaction region is much larger than the Fermi
length of electrons in  metallic grains.
Therefore we can change quantum numbers ($mln$) to $(k_{x}k_y k_z)$. Introducing $k_\perp=\sqrt{k^2_x+k^2_y}\approx l/a$ and
the total electron energy, $E^s_\bi k=U+ s\Ji+\hbar^2(k^2_\perp+k_z^2)/(2m_\mathrm e)$
we obtain the electron wave functions outside and inside the grains in
(\ref{Eq:WaveFuncR5}) and (\ref{Eq:WaveFuncR6}).

\section{Matrix elements}\label{MatElApp}

\subsection{FM ordering}

Using (\ref{Eq:WaveFuncR5}) and (\ref{Eq:WaveFuncR6}) we find for matrix elements
the following results
\begin{equation}\label{Eq:ME2}
\eqalign{
 V^s_{ii}=\int_{-\tilde\lambda_\perp}^{\tilde\lambda_\perp} dxdy \int_{d/2}^\infty dz|\psi^s_i|^2(-U+s\Ji)= \cr
 =(s\!\Ji-U)\frac{(|\tau^s_{i}|)^2}{2\Omega\varkappa_i^s}\rme^{-2\varkappa_i^s d} F_1(q),\cr
 P_{ij}^s=\int_{-\tilde\lambda_\perp}^{\tilde\lambda_\perp} dxdy \int_{-\infty}^\infty dz\phi^{s*}_i\psi^s_j=\frac{  T_{ij}^s+ T_{ji}^{s*}}{s\!\Ji-U}+\cr +\frac{\tau^{s*}_{i}\tau^s_j \rme^{-(\!\varkappa^s_i+\varkappa^s_j\!)\frac{d}{2}}\mathrm{sinh}(\!(\!\varkappa^s_i-\varkappa^s_j\!)\frac{d}{2})}{\Omega(\varkappa_i^s-\varkappa^s_j)}F_2(q),\cr
   T^s_{ij}=\int_{-\tilde\lambda_\perp}^{\tilde\lambda_\perp} dxdy \int_{d/2}^\infty dz\phi^{s*}_i\psi^s_j(-U+s\Ji)=\cr =(s\!\Ji-U)\frac{\tau^{s*}_{i}\tau^s_{j}(\varkappa_i^s+\varkappa_j^s)}{\Omega((k_i^s)^2+(\varkappa_j^s)^2)}\rme^{-\varkappa_j ^s d}F_3(q).
}
\end{equation}
The wave functions are overlapped in a finite area in the (x,y)-plane and
therefore $k_x$ and $k_y$ components of electron momentum do not
conserve during the transitions. To calculate matrix
elements $V_{ii}^s$, $P_{ij}^s$ and $T_{ij}^s$ we approximate the
wave functions in (\ref{Eq:WaveFuncR5}) and (\ref{Eq:WaveFuncR6}) with
plane waves confined within a square window in the perpendicular direction.
The size of this window is defined by the suppression factor $\rme^{-\varkappa r^2_\perp/2a}$
in the expression for the wave functions. We change this factor with
a step function being finite in a square $|x|<\tilde \lambda_\perp$, $|y|<\tilde \lambda_\perp$ and zero outside this region.
Therefore, instead of circular integral region we consider the rectangular region.
Integrals of the type $\int_{|x|<\lambda_\perp}\rme^{-\rmi q_x x}\sim \mathrm{Sinc}(q\lambda_\perp)$
produce $\mathrm{Sinc}$-like factors in the matrix elements. The size of
$\tilde \lambda_\perp$ is different for different matrix elements and is defined by the wave functions overlap
area. For matrix elements $V^s_{ii}$ the size of the overlap area is $\lambda_\perp^2= \pi a/\varkappa_i$, for the tunneling matrix elements $T^s_{ij}$ the size is $\lambda_\perp^2=2\pi a/\varkappa_j$. The overlap term $P^s_{ij}$ has three contributions: The first two have the same
area as $T^s_{ij}$ and the last contribution has
the area $\lambda_\perp^2=2\pi a/(\varkappa_i+\varkappa_j)$. Using these approximations
we have the following result for functions $F_i$ in (\ref{Eq:ME2})
\begin{equation}\label{Eq:MEPerp}
\eqalign{
 F_1=\frac{\pi a}{\varkappa^s_i}\mathrm{sinc}\left(\frac{q_x}{2}\sqrt{\frac{\pi a}{\varkappa^s_i}}\right)\mathrm{sinc}\left(\frac{q_y}{2}\sqrt{\frac{\pi a}{\varkappa^s_i}}\right),\cr
 F_2=\frac{2\pi a}{\varkappa^s_i+\varkappa^s_j}\mathrm{sinc}\!\!\left(\frac{q_x}{2}\sqrt{\frac{2\pi a}{\varkappa^s_i+\varkappa^s_j}}\right)\!\!\mathrm{sinc}\!\!\left(\frac{q_y}{2}\sqrt{\frac{2\pi a}{\varkappa^s_i+\varkappa^s_j}}\right),\cr
 F_3=\frac{2\pi a}{\varkappa^s_j}\mathrm{sinc}\left(\frac{q_x}{2}\sqrt{\frac{2\pi a}{\varkappa^s_j}}\right)\mathrm{sinc}\left(\frac{q_y}{2}\sqrt{\frac{2\pi a}{\varkappa^s_j}}\right),
}
\end{equation}
where $\bi q=\bi k_{i\perp}-\bi k_{j\perp}$, $k_{i\perp}=(k_{ix},k_{iy},0)$ and $q=|\bi q|$.
Factors $F_{1,2,3}$ decay rapidly outside the region $|q|>1/\lambda_\perp$.

\subsection{AFM ordering}

For AFM ordering the grain magnetic moments are anti-parallel.
The superscripts in all matrix elements refer to the spin state in the first grain.
The spin state in the second grain is the opposite to the spin state
in the first grain. Using the same approach as above for matrix elements we find
\begin{equation}\label{Eq:ME7}
\eqalign{
 V^s_{ii}=(-s\!\Ji-U)\frac{(|\tau^s_{i}|)^2}{2\Omega(\varkappa_i^s)}\rme^{-2\varkappa_i^s d}F_1(q),\cr
   T^s_{ij}\!=(-s\!\Ji-U)\frac{\tau^{-s*}_{i}\tau^s_{j}(\varkappa_i^{-s}\!\!+\varkappa_j^s)}{\Omega(\!(k_i^{-s})^2+\!(\varkappa_j^{s})^2)}\rme^{-\varkappa_j^{s} \!d}F_3(q),\cr
 P_{ij}^s=\frac{T_{ji}^{-s*}}{s\!\Ji-U}+\frac{  T_{ij}^{s}}{-s\!\Ji-U}+\\ +\frac{ \tau^{-s*}_{i}\tau^s_j \rme^{-(\!\varkappa^{-s}_i+\varkappa^s_j\!)\frac{d}{2}}\mathrm{sinh}(\!(\!\varkappa_i^{-s}-\varkappa_j^{s}\!)\frac{d}{2})}{\Omega((\varkappa_i^{-s})-(\varkappa_j^s))}F_4(q),
}
\end{equation}
where
\begin{equation}\label{Eq:MEPerp1}
F_4=\frac{2\pi a}{\varkappa^{-s}_i\!\!+\!\varkappa^s_j}\mathrm{sinc}\!\!\left(\!\frac{q_x}{2}\sqrt{\frac{2\pi a}{\varkappa^{-s}_i\!\!+\!\varkappa^s_j}}\right)\!\!\mathrm{sinc}\!\!\left(\!\frac{q_y}{2}\sqrt{\frac{2\pi a}{\varkappa^{-s}_i\!\!+\!\varkappa^s_j}}\right)\!.
\end{equation}

\section{Formalism}\label{Formalism}

Here we consider only the case of FM ordering of grains magnetic moments. The case of AFM ordering can be considered in a similar way. The zero order wave function is the Slater determinant
\begin{equation}\label{Eq:ZeroOrderWF}
\Psi_0=\frac{1}{\sqrt{N}}\left(\eqalign{ \psi^{s_1}_1(\bi r_1) ~~~  ... ~~~  \psi^{s_1}_1(\bi r_{2n_0})\cr
~~~~~~~~~~~~~...  \cr
\phi^{s_{n_0+1}}_1(\bi r_1)~ ... ~ \phi^{s_{n_0+1}}_1(\bi r_{2n_0})\cr
 ~~~~~~~~~~~~~...  }\right).
\end{equation}
States $\psi_i$ and $\phi_j$ are chosen such that all the
energy levels below $E_\mathrm F$ are filled: $n_0$ states in the
left grain and $n_0$ states in the right grain.
The normalizing factor is
\begin{equation}\label{Eq:NormFact}
N=(2n_0)!(1-2\sum_{i,j,s}|P_{ij}^s|^2+...),
\end{equation}
where $i$ and $j$ enumerate states in the left and the right grains, respectively.
The second term in (\ref{Eq:NormFact}) appears due to nonorthogonality of the basis wave functions.
Further we introduce the excited states as follows
\begin{equation}\label{Eq:ExcWF}
\Psi_{ij}^s=\hat b^{s+}_i \hat a^s_j \Psi_0 = \frac{1}{\sqrt{N_{ij}}}\left(\eqalign{ \psi^{s_1}_1(\bi r_1) ~~~~~  ... ~~~~~  \psi^{s_1}(\bi r_{2n_0})\cr
 ~~~~~~~~~~~~~~...  \cr
\phi^{s_{n_0+j-1}}_{j-1}(\bi r_1)~ ... ~ \phi^{s_{n_0+j-1}}_{j-1}(\bi r_{2n_0})\cr
\psi^{s}_{i}(\bi r_1)~~~~~~ ... ~~~~~~ \psi^{s}_{i}(\bi r_{2n_0})\cr
\phi^{s_{n_0+j+1}}_{j+1}(\bi r_1) ~... ~ \phi^{s_{n_0+j+1}}_{j+1}(\bi r_{2n_0})\cr
 ~~~~~~~~~~~~~~...  }\right).
\end{equation}
The annihilation operator removes a line in the Slater determinant while the
creation operator adds a line. The normalization factor has the form $N_{ij}=(2n_0)!(1-2\sum_{k,l,s}|P_{kl}^s|^2+2\sum_{k,s}(2\sum_{k,s}(|P_{kj}^s|^2+|P_{jk}^s|^2)-2\sum_{k,s}(2\sum_{k,s}(|P_{ki}^s|^2+|P_{ik}^s|^2)...)$.
One can introduce the excited wave function $\tilde \Psi^s_{ij}=\hat a^{s+}_i \hat b^s \Psi_0$.
These wave functions correspond to single excitations with only one electron transferred between
grains.  The Coulomb energy of excited states is $\tilde \epsilon_\mathrm c$.  In our calculations
we neglect states with two electrons transferred between grains.
Such states have large Coulomb energy, $2\tilde \epsilon_\mathrm c$ and
therefore transitions to these states have much lower probability.
Also we neglect electron transitions between the single particle states within
the same grain ($\psi_i\to\psi_j$).
The probability of such transitions is of order $p^2$.
Therefore, these transitions contribute to the system energy on the level of $p^4$ which is beyond
our accuracy ($p^2$). Using the above excited states we can write the
perturbed wave function in (\ref{Eq:WaveFuncFM}).

To find the coefficients $\beta^s_{ij}$ and $\tilde \beta^s_{ij}$ in (\ref{Eq:WaveFuncFM})
we solve the Schrodinger equation
\begin{equation}\label{Eq:ShrodEq}
(\hat H_{\mathrm{sp}}+\hat H_\mathrm C-E)|(1+\alpha_0)\Psi_0+\!\!\!\sum_{s,i,j} \beta_{ij}^s\Psi^s_{ij}+\!\!\!\sum_{s,i,j} \tilde\beta_{ij}^s\tilde\Psi^s_{ij}\rangle=0.
\end{equation}
Selecting terms of order of $p$ we find
\begin{equation}\label{Eq:ShrodEq1}
\beta^s_{ij}=\tilde \beta^s_{ij}=-\frac{\langle\Psi^s_{ij}|(\hat H_{\mathrm{sp}}+\hat H_\mathrm C-E)|\Psi_0\rangle}{\langle\Psi^s_{ij}|(\hat H_{\mathrm{sp}}+\hat H_\mathrm C-E)|\Psi^s_{ij}\rangle}.
\end{equation}
We neglect electron transitions between the grains due to
the Coulomb interaction,
$\langle\Psi^s_{ij}|\hat H_\mathrm C|\Psi_0\rangle=0$. Using (\ref{Eq:ZeroOrderWF}) and
(\ref{Eq:ExcWF}) we find
\begin{equation}\label{Eq:Trans}
\langle\Psi^s_{ij}|\hat H_{\mathrm{sp}}|\Psi_0\rangle=T^s_{ij}+E_0 P^s_{ij}\sim p,
\end{equation}
where $E_0=2\sum_{i,s}\epsilon_i^s$ (the summation is over
states below the Fermi energy). Finally we obtain
\begin{equation}\label{Eq:Trans1}
\langle\Psi^s_{ij}|(\hat H_{\mathrm{sp}}-E)|\Psi_0\rangle=T^s_{ij},
\end{equation}
here we take into account the fact that the
total energy is $E=E_0+O(p^2)$. The denominator in (\ref{Eq:ShrodEq1})
is given by the expression
\begin{equation}\label{Eq:Trans2}
\langle\Psi^s_{ij}|\hat H_{\mathrm{sp}}|\Psi_{ij}^s\rangle=\epsilon^s_{i}-\epsilon^s_j+\tilde \epsilon_\mathrm c.
\end{equation}
Using (\ref{Eq:Trans1}) and (\ref{Eq:Trans2}) we obtain (\ref{Eq:WaveFuncFM1})
in the main text.

To find the coefficient $\alpha_0$ in (\ref{Eq:WaveFuncFM})
we use the fact that the total number of electrons
in the system is conserved. Introducing the operator of the total number of
electrons, $\hat N=\sum_{i,s}\hat a^{s+}_i\hat a^s_i+\sum_{i,s}\hat b^{s+}_i\hat b^s_i$ we have the following relation
\begin{equation}\label{Eq:PartNumbFM}
\eqalign{
 1=\frac{\langle\Psi|\hat N|\Psi\rangle}{N_0}=(1+2\alpha_0)-4\!\!\!\!\!\!\!\!\!\sum_{s,i\notin S_0^s,j\in S_0^s}\!\!\frac{\mathrm{Re}(  T^s_{ij} P^{s*}_{ij})}{\epsilon_i^s-\epsilon_j^s+\tilde \epsilon_\mathrm c}+\cr
 +2\!\!\!\!\!\!\!\!\!\sum_{s,i\notin S_0^s,j\in S_0^s}\!\!\frac{|  T^s_{ij}|^2}{(\epsilon_i^s-\epsilon_j^s+\tilde \epsilon_\mathrm c)^2}.
}
\end{equation}
Now we can calculate the system energy
\begin{equation}\label{Eq:PartNumbFM}
\eqalign{
 E^\mathrm{FM}=\langle\Psi|\hat H_\mathrm{sp}+\hat H_\mathrm C|\Psi\rangle=(1+2\alpha_0)\langle\Psi_0|\hat H_\mathrm{sp}+\hat H_\mathrm C|\Psi_0\rangle+\cr
 +4\sum_{i,j,s}\mathrm{Re}\left(\beta^{s*}_{ij}\langle\Psi^s_{ij}|\hat H_\mathrm{sp}+\hat H_\mathrm C|\Psi_0\rangle\right)+\cr
 +2\sum_{i,j,s}\mathrm{Re}\left(|\beta^{s}_{ij}|^2\langle\Psi^s_{ij}|\hat H_\mathrm{sp}+\hat H_\mathrm C|\Psi^s_{ij}\rangle\right).
}
\end{equation}
When calculating the first term, $\langle\Psi_0|\hat H_\mathrm{sp}+\hat H_\mathrm C|\Psi_0\rangle$,
the corrections of the order of $p^2$ in the normalization factor $N$ should be
taken into account. We obtain the following formula for the energy of the FM state
\begin{equation}\label{Eq:EnFM}
\eqalign{
E^\mathrm{FM}= 2\!\!\!\sum_{s,i\in S_0^s}(\epsilon_i^s+V^s_{ii})-2\!\!\!\!\!\sum_{s,i\in S_0^s, j\in S_0^s}  \mathrm Re(T^s_{ij}P^{s*}_{ij})-\cr
 -2\!\!\!\!\!\!\!\!\!\sum_{s,i\notin S_0^s, j\in S_0^s}\frac{|  T^s_{ij}|^2}{\epsilon^s_i-\epsilon^s_j+\tilde \epsilon_\mathrm c}.
}
\end{equation}
We use the fact that the mean energy level spacing is much smaller than the Fermi energy and
replace the summation with integration in (\ref{Eq:EnFM}). We introduce momentum $\bi k$ instead of
orbital numbers $i$ and replace the integration over the $\bi k_{1,2\perp}$ with integration over the
$\bi q=\bi k_{1\perp}-\bi k_{2\perp}$ and $\tilde{\bi k}=(1/2)(\bi k_{1\perp}+\bi k_{2\perp})$.
In general the boundaries of integration in these new coordinates are rather
complicated. However, as we showed in the previous section all
matrix elements depend only on $\bi q$ and not on $\tilde{\bi k}$.
In addition, all matrix elements are finite only in the small region,
$|\bi q|<\pi/\lambda_\perp$. Thus, we can integrate over $\bi q$ independently
of $\tilde{\bi k}$ in the whole $\bi k_\perp$-space. Taking this into account
we obtain (\ref{Eq:EnFM1}) describing the energy of the FM state.
The case of AFM configuration of $\bi M_1$ and $\bi M_2$ can be considered in a similar way.

\section*{References}

\bibliography{Exchange}

\end{document}